\documentclass[a4paper,fleqn,usenatbib]{mnras}

\usepackage{newtxtext,newtxmath}

\usepackage[T1]{fontenc}
\usepackage{ae,aecompl}

\usepackage{graphicx}	
\usepackage{amsmath}	
\usepackage{amssymb}	
\usepackage[normalem]{ulem}

\title[MPs within Mayall\,II in M31]{{\it Hubble Space Telescope} analysis of stellar populations within the globular cluster G1 (Mayall\,II) in M\,31\thanks{Based on observations with the NASA/ESA {\it Hubble
      Space Telescope}, obtained at the Space Telescope Science
    Institute, which is operated by AURA, Inc., under NASA contract
    NAS 5-26555.}}

\author[Nardiello et al.]{D.\ Nardiello$^{1,2}$\thanks{E-mail: domenico.nardiello@unipd.it},
G.\ Piotto$^{1,2}$,
A.\ P.\ Milone$^1$,
R.\ M.\ Rich$^3$,
S.\ Cassisi$^{4,5}$,
L.\ R.\ Bedin$^2$,
\newauthor
A.\ Bellini$^6$,
A.\ Renzini$^2$ \\
$^{1}$Dipartimento di Fisica e Astronomia ``Galileo Galilei'', Universit\`a di Padova, Vicolo dell'Osservatorio 3, Padova IT-35122 \\
$^{2}$Istituto Nazionale di Astrofisica - Osservatorio Astronomico di Padova, Vicolo dell'Osservatorio 5, Padova, IT-35122 \\
$^{3}$Department of Physics and Astronomy, UCLA, 430 Portola Plaza, Box 951547, Los Angeles, CA 90095-1547, USA \\
$^{4}$INAF -- Osservatorio Astronomico d'Abruzzo, Via M. Maggini, I-64100 Teramo, Italy \\
$^{5}$INFN -- Sezione di Pisa, Largo Pontecorvo 3, I-56127 Pisa, Italy \\
$^{6}$Space Telescope Science Institute, 3800 San Martin Drive, Baltimore, MD 21218, USA \\
}

\date{Accepted 2019 March 1. Received 2019 February 25; in original form 2018 December 5}

\pubyear{2018}

\begin{document}
\label{firstpage}
\pagerange{\pageref{firstpage}--\pageref{lastpage}}
\maketitle

\begin{abstract}

In this paper we present a multi-wavelength analysis of the complex
stellar populations within the massive globular cluster
Mayall\,II (G1), a satellite of the nearby Andromeda galaxy projected
at a distance of 40\,kpc.  We used images collected with the {\it
  Hubble Space Telescope} in UV, blue and optical filters to explore
the multiple stellar populations hosted by G1.

The $m_{\rm F438W}$ versus $m_{\rm F438W}- m_{\rm F606W}$
colour-magnitude diagram shows a significant spread of the red giant
branch, that divides $\sim 1$\,mag brighter than the red clump. A
possible explanation is the presence of two populations with different
iron abundance  or different
  C+N+O content, or different helium content, or a combination of the
  three causes. A similar red giant branch split is observed also for
the Galactic globular cluster NGC\,6388.

Our multi-wavelength analysis gives also the definitive proof that G1
hosts stars located on an extended blue horizontal branch.  The horizontal branch of
G1 exhibits similar morphology as those of NGC\,6388 and NGC\,6441,
which host stellar populations with extreme helium abundance
(Y>0.33). As a consequence, we suggest that G1 may also exhibit large
star-to-star helium variations.

\end{abstract}




\section{Introduction}

 The {\it Hubble Space Telescope} ({\it HST}) Ultraviolet Legacy
 Survey of Galactic Globular Clusters (GCs, GO-13297, PI:~Piotto) has
 identified and characterised the multiple stellar populations (MPs)
 from homogeneous multi-band photometry of a large sample of 58 GCs.
 All GCs host two or more distinct populations and the complexity of
 the multiple-population phenomenon increases with the cluster
 mass. Specifically, the MPs of the most massive Galactic GCs exhibit
 extreme chemical compositions -- such as large helium and
   light-(and in some cases also heavy-)elements variations among the
   cluster members -- and their present population is dominated by
   second-generation stars born from matter polluted by stars
   belonging to the first generation; these properties appear more
   extreme than those commonly observed in less massive GCs of the
   Milky Way  (\citealt{2015AJ....149...91P,2017MNRAS.464.3636M}).

\begin{figure*}
\includegraphics[width=0.80\textwidth]{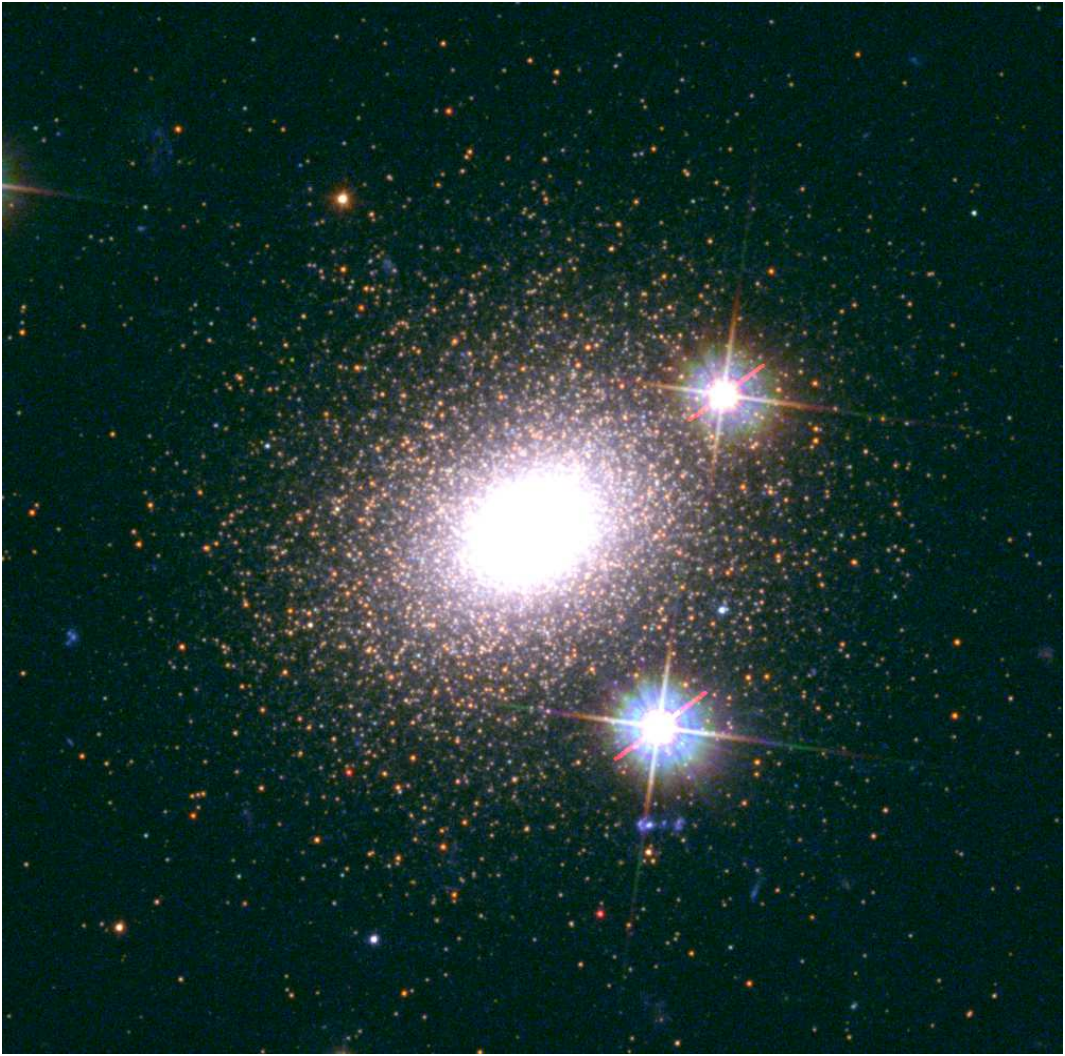}
\caption{Stacked three-colour image (F336W$+$F438W$+$F606W) of a field
  39.5$\times$39.5\,arcsec$^2$ centred on G1. North is up, east is
  left. \label{fig:1}}

\end{figure*}
\begin{table*}
  \caption{Description of the archival {\it HST} images reduced in this analysis.}
    \label{tab1}
    \begin{tabular}{ l c c c c c}
\hline
\multicolumn{6}{c}{MAYALL\,II (G1)} \\
\hline
\multicolumn{1}{c}{Program} &
\multicolumn{1}{c}{Epoch} &
\multicolumn{1}{c}{Filter} &
\multicolumn{1}{c}{$N \times $ Exp.~time} &
\multicolumn{1}{c}{Instrument} &
\multicolumn{1}{c}{PI} \\ 
\hline
5464  &  1994.58 & F555W &   4 $\times$ 400\,s & WFPC2 &  Rich \\
5464  &  1994.58 & F814W &   3 $\times$ 400\,s & WFPC2 &  Rich \\
5907  &  1995.76 & F555W &   2 $\times$ 500\,s $+$ 2 $\times$ 600\,s & WFPC2 &  Jablonka \\
5907  &  1995.76 & F814W &   2 $\times$ 400\,s $+$ 2 $\times$ 500\,s & WFPC2 &  Jablonka \\
9767  &  2003.81 & F555W &   6 $\times$ 410\,s & ACS/HRC &  Gebhardt \\
12226 &  2011.01 & F275W &   2 $\times$1395\,s $+$ 4 $\times$1465\,s & WFC3/UVIS &  Rich \\
12226 &  2010.85 & F336W &   6 $\times$ 889\,s                       & WFC3/UVIS &  Rich \\
12226 &  2011.01 & F438W &   2 $\times$1395\,s $+$ 4 $\times$1465\,s & WFC3/UVIS &  Rich \\
12226 &  2010.85 & F606W &    6 $\times$ 889\,s & WFC3/UVIS &  Rich \\
\hline
\end{tabular}

\end{table*}

\begin{figure*}
\includegraphics[width=0.99\textwidth, bb=18 181 583 585]{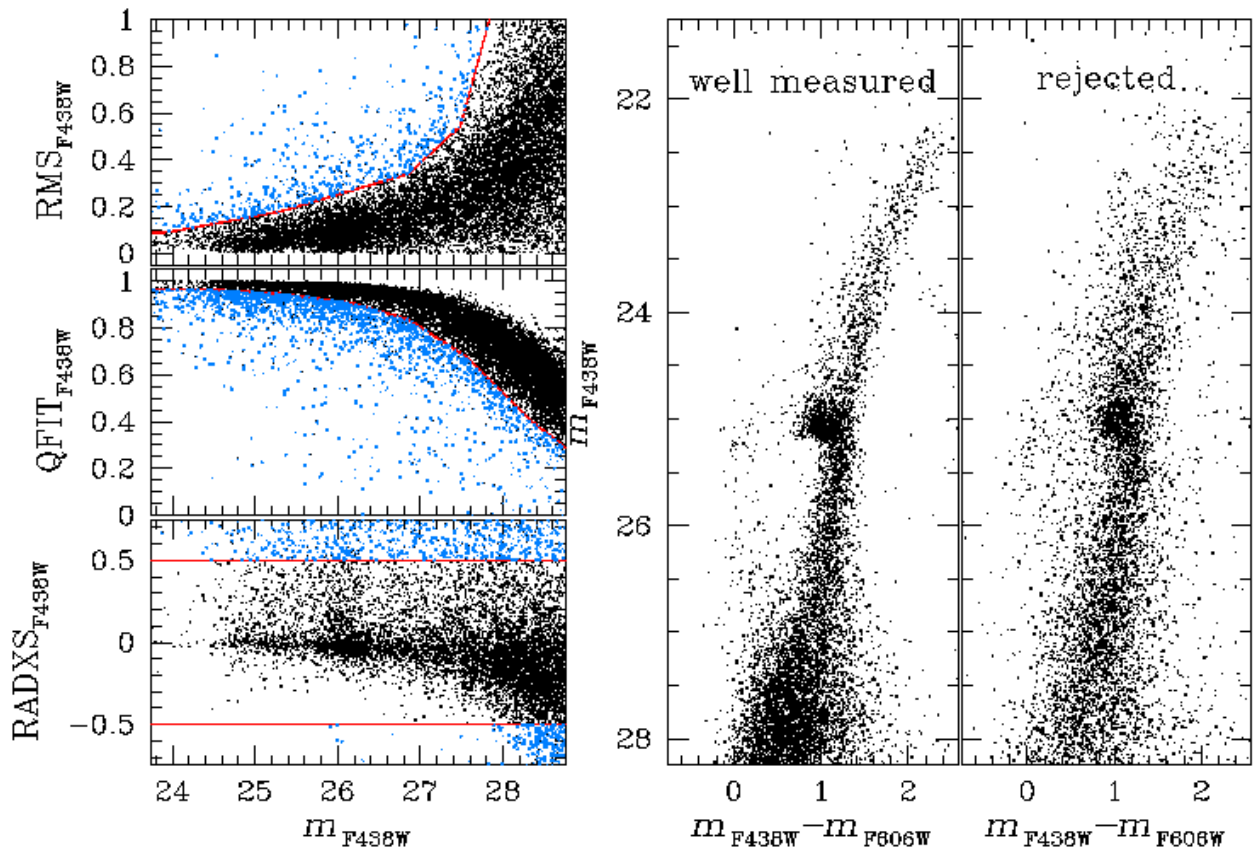}
\caption{Selection of well-measured stars for G1. Left-hand
    panels show the selections (black points) for the filter F438W
    using the parameters \texttt{RMS} (top panel), \texttt{QFIT}
    (center panel), and \texttt{RADXS} (bottom panel). Middle and
    right panels show the $m_{\rm F606W}$ versus $m_{\rm
      F438W}-m_{F606W}$ CMD for the stars that pass the selection
    criteria and the rejected stars, respectively.\label{fig:1bis}}
\end{figure*}

\begin{figure*}
\includegraphics[width=0.95\textwidth, bb=15 275 510 575]{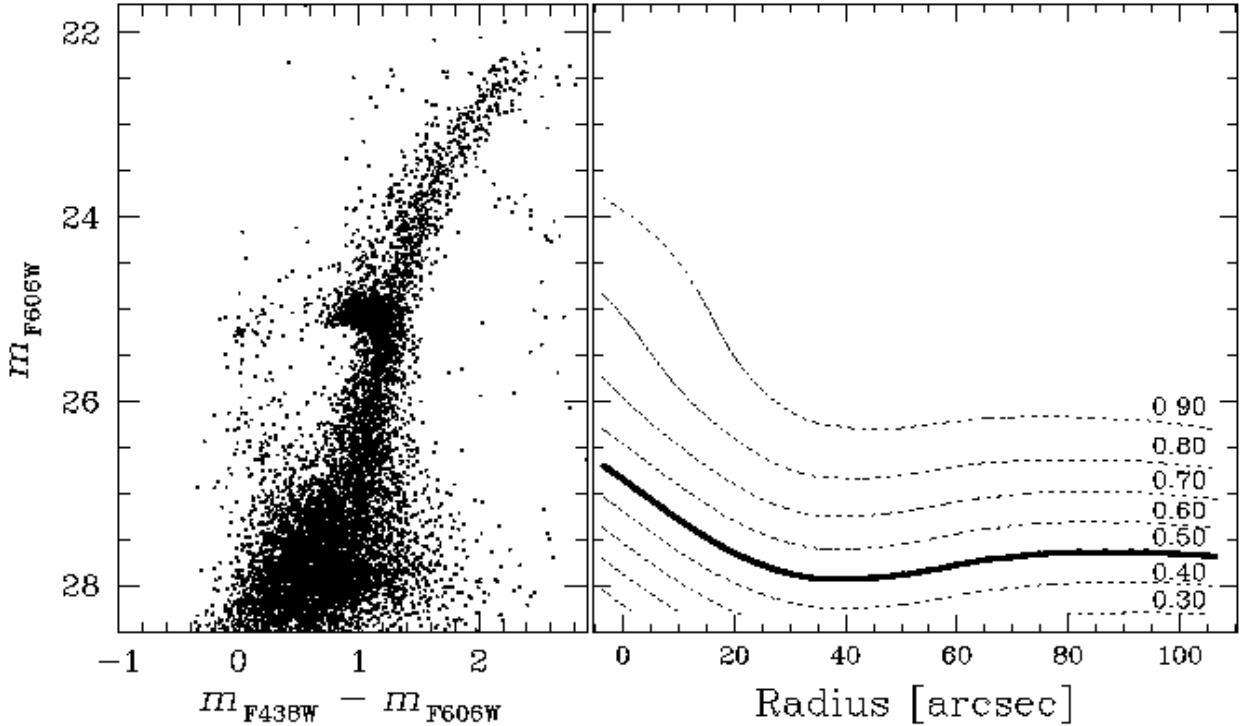}
\caption{Left panel shows the $m_{\rm F606W}$ versus $m_{\rm
    F438W}-m_{\rm F606W}$ CMD of all the well measured stars in the
  field of view containing G1. Right panel shows the completeness
  levels of our catalogue as a function of the F606W magnitude and of
  the radial distance from the centre of the cluster. \label{fig:2}}
\end{figure*}

\begin{figure}
\includegraphics[width=0.49\textwidth, bb=18 144 422 696]{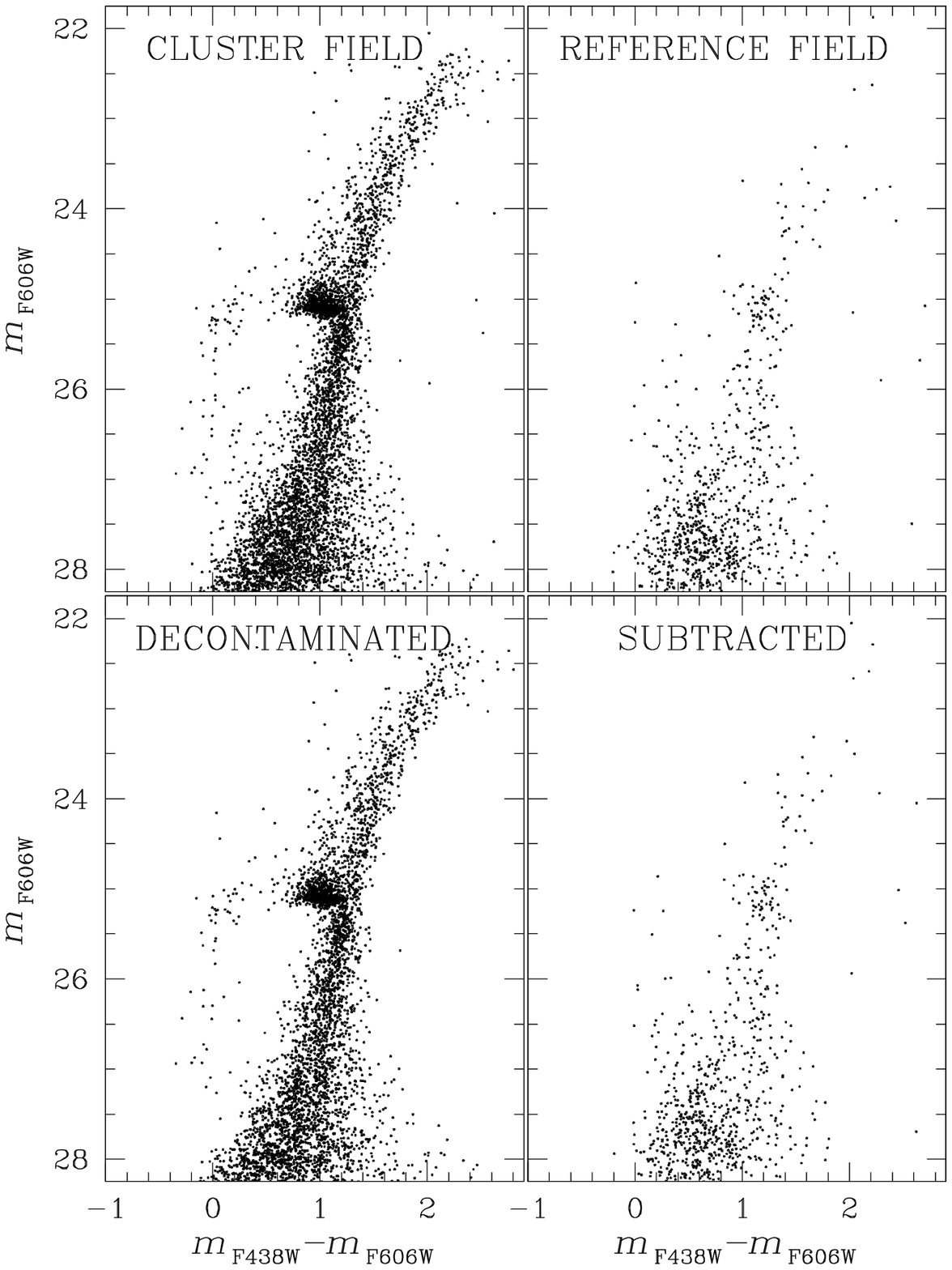}
\caption{Statistical removal of field stars. Top panels show the
  $m_{\rm F606W}$ versus $m_{\rm F438W}-m_{\rm F606W}$ CMDs of the
  cluster (left) and reference (right) fields. Bottom-left panel shows
  the $m_{\rm F606W}$ versus $m_{\rm F438W}-m_{\rm F606W}$ CMD of 
  cluster-field stars after the decontamination; bottom-right panel
  shows the stars subtracted to the CMD of the cluster
  field.  \label{fig:3}}
\end{figure}

In the majority of the analysed GCs, MPs have uniform metallicity and
are characterised only by different helium content and light-element
abundance. However, some massive clusters also show internal
variations in heavy elements, including iron and s-process elements
(e.g.\,\citealt{2015MNRAS.450..815M,2015AJ....150...63J}).  In particular, $\omega$\,Cen,
which is the most-massive Milky-Way GC ($\mathcal{M}= 10^{6.4}
\mathcal{M}_{\rm \odot}$, \citealt{2005ApJS..161..304M}), exhibits the
most-extreme variation of iron, with [Fe/H] ranging from $\sim -2.2$
to $\sim -0.7$
(e.g.\,\citealt{1996ASPC...92..375N,2010ApJ...722.1373J,2011ApJ...731...64M}).

Understanding the properties of MPs in GCs may constrain the models of
formation of the GCs in the early Universe, the mechanisms responsible
for the assembly of the halo, and the role of GCs in the re-ionization
of the Universe (e.g., \citealt{2015MNRAS.454.4197R,2017MNRAS.469L..63R}
and references therein).

Evidence of MPs in GCs belonging to external galaxies are widely reported in
  literature.  Photometric detection of MPs in four GCs of the Fornax
  dSph have been reported by \citet{2014ApJ...797...15L}. Multiple
  sequences have also been observed in the CMDs of young and
  intermediate age clusters of the Small and Large Magellanic Clouds
  (see, e.g.,
  \citealt{2017MNRAS.464...94N,2017MNRAS.465.4159N,2017MNRAS.467.3628C,2018MNRAS.tmp..637M,2017MNRAS.465.4363M}
  and references therein), but the phenomenon that generates these
  multiple sequences may be different from that observed for old stellar
  systems. Spectroscopic detection of MPs in GCs of external galaxies
  have been reported, e.g., by \citet[in the Fornax dSph
    ]{2012A&A...546A..53L}, \citet[in
    M\,31]{2013ApJ...776L...7S}, \citet[in M81]{2013MNRAS.436.2763M},
  \citet{2009ApJ...695L.134M}, \citet{2016ApJ...829...77D}, and
  \citet[in the Large and Small Magellanic
    Clouds]{2018MNRAS.477.4696M}. 

In this context, the investigation of massive clusters in nearby
galaxies is of great importance to understand whether the occurrence
of MPs and the dependence on the mass of the host cluster are a
peculiarity of Milky-Way GCs or are an universal phenomenon.

To address this issue, we investigate the presence of MPs in the GC
Mayall II (G1, \citealt{1977AJ.....82..947S}), in the nearby galaxy
M\,31, a cluster three times more massive than $\omega$\,Cen
($\mathcal{M}=10^{7.2} \mathcal{M}_{\rm \odot}$,
\citealt{2001AJ....122..830M}).  G1 has been previously investigated
by using Wide Field and Planetary Camera 2 photometry in the F555W and
F814W bands
(\citealt{1996AJ....111..768R,2001AJ....122..830M,2007MNRAS.376.1621M,2012A&A...544A.155F}).
Early evidence of metallicity variations is provided by
\citet{2001AJ....122..830M} on the basis of the broadened RGB.  It is
noteworthy that G1 
exhibits kinematic hints for a central intermediate mass
black hole (\citealt{2005ApJ...634.1093G}).

In this work we present exquisite multi-band photometry of the stars
hosted by G1, extracting their luminosities in six {\it HST} filters,
from UV to optical bands (Section~\ref{sec:obs}).  We identify major features
of the CMD of G1 never shown before and we analyse them to extract
the main properties of the cluster stellar populations
(Section~\ref{sec:cmd}). Finally, we compare the results obtained from
G1 with those for massive Galactic GCs that present similar features in
the CMD (Section~\ref{sec:compare}), and we discuss them
(Section~\ref{sec:summary}) in the context of multiple stellar
populations.

\section{Observations and data reduction}
\label{sec:obs}

For this work we reduced the archival {\it HST} data of G1 collected
with the Wide Field and Planetary Camera 2 (WFPC2), with the
High-Resolution Channnel (HRC) of the Advanced Camera for Surveys
(ACS), and with the UVIS imager of the Wide Field Camera 3 (WFC3).
WFPC2 data in F555W and F814W bands were collected within programs
GO-5464 (PI:~Rich) and GO-5907 (PI:~Jablonka); observations in F555W
with ACS/HRC were obtained within GO-9767 (PI:~Gebhardt); WFC3/UVIS
images in F275W, F336W, F438W, and F606W bands were taken for GO-12226
(PI:~Rich). Table~\ref{tab1} gives the journal of observations used in
this work. Figure~\ref{fig:1} shows a stacked three-colour image of the G1
field.

To obtain high precision photometry, we perturbed library point spread
functions\footnote{http://www.stsci.edu/$\sim$jayander/STDPSFs} (PSFs)
to extract, for each image, a spatial- and time-varying array of PSFs,
as already done in other works by our group (see, e.g.,
\citealt{2017ApJ...842....6B,2018MNRAS.tmp..637M,2018MNRAS.tmp..706N,2018MNRAS.481.3382N}).
These PSF arrays have been used to extract positions and fluxes of the
stars on the images. We corrected the positions using the geometric
distortion solutions given by \citet[WFPC2]{2003PASP..115..113A},
\citet[ACS/HRC]{2004acs..rept....3A}, \citet{2009PASP..121.1419B} and
\citet[WFC3/UVIS]{2011PASP..123..622B}. We used the Gaia DR2 catalogue
(\citealt{2016A&A...595A...2G,2018arXiv180409365G}) to transform the
positions of the single catalogues in a common reference frame.  We
used the \texttt{FORTRAN} routine \texttt{kitchen\_sync2} (KS2, for
details see
\citealt{2008AJ....135.2055A,2017ApJ...842....6B,2018MNRAS.481.3382N})
that takes as input the PSF arrays, the transformations and the
images, and analyses all the images simultaneously to find the
position and measure the flux of each source, after subtracting its
neighbours. This routine allows us to obtain high precision photometry
in crowded environments and of stars in the faint magnitude regime.
We calibrated the output photometry into the Vega-mag system using,
for WFPC2, the zero-points and the aperture corrections tabulated by
\citet{1995PASP..107.1065H}, for ACS/HRC the zero-points given by the
``ACS Zeropoints
Calculator''\footnote{https://acszeropoints.stsci.edu} and the
aperture corrections tabulated by \citet{2016AJ....152...60B}, and for
WFC3/UVIS the zero-points and the aperture corrections listed by
\citet{2017wfc..rept...14D}.

To perform our analysis, we selected only well measured stars as in
\citet{2018MNRAS.tmp..706N} on the basis of several diagnostic
parameters output of the routine KS2 (see
\citealt{2018MNRAS.481.3382N} for details). The procedure is
  illustrated in Fig.~\ref{fig:1bis} for the filter F438W, but it is
  the same for all the other filters. Briefly, we divided the
  distributions of the parameters \texttt{RMS} (the standard deviation
  of the mean magnitude divided $\sqrt{N-1}$ with $N$ the number of
  measurements) and \texttt{QFIT} (the quality-of-fit) in intervals of
  0.5 magnitude and, in each magnitude bin, we calculated the
  3.5$\sigma$-clipped average of the parameter and then the point
  3.5$\sigma$ above the mean value of the parameter. We linearly
  interpolated these points (red line) and we excluded all the points
  above (in the case of the \texttt{RMS}) or below (for the
  \texttt{QFIT}) this line (azure points). Moreover, we excluded all
  the sources that have \texttt{RADXS}$>0.5$ (where \texttt{RADXS} is
  the shape parameter, defined in \citealt{2008ApJ...678.1279B}) and
  \texttt{RADXS}$<-0.5$ (bottom left-hand panel of
  Fig.~\ref{fig:1bis}), and for which the number of measurements is
  $N<2$. Middle-panel of Fig.~\ref{fig:1bis} shows the $m_{\rm F606W}$
  versus $m_{\rm F438W}-m_{F606W}$ CMD for the stars that pass the
  selections in F438W and F606W filters, while the right-hand panel
  shows the rejected stars.

\subsection{Artificial stars and completeness}

We estimated the completeness of our G1 catalogue using artificial
stars (AS). We produced 100\,000 ASs having magnitude $21.75 \le
m_{\rm F606W} \le 28.25$, flat luminosity function in F606W, colours
that lie along the red giant branch (RGB) and the horizontal branch
(HB) fiducial lines, and flat spatial distribution.  For each AS in
the input list, the routine KS2 added the star to each image, searched
and measured it using the same approach used for real stars, and then
removed the source and passed to the next AS (see
  \citealt{2008AJ....135.2055A} for details). Using this approach, if
  an artificial star is added on a crowded zone of the image (like the
  cluster center), comparing the input and output photometry of the
  artificial star we can estimate how much the crowding is affecting
  the measurement of a star located in that position.

To estimate the level of completeness, we considered an AS as
recovered if the difference between the input and output positions
is $<0.5$ pixel and the difference between the input and output
$m_{\rm F606W}$ and $m_{\rm F438W}$ magnitudes is $<0.75$ mag.

Figure~\ref{fig:2} illustrates the completeness of our catalogue: right
panel shows the completeness contours in the $m_{\rm F606W}$ versus
radial distance plane. Our catalogue has completeness $>50$\% for
stars with $m_{\rm F606W}\lesssim 26.5$ at all radial distances from the
cluster centre.

\subsection{Field decontamination}

Due to the large distance of G1 and the short baselines of the
observations, using proper motions to separate G1 members from field
stars is not possible.  
We decontaminated the cluster CMD from background and foreground stars
using the statistical approach (see, for example,
\citealt{2018MNRAS.tmp..637M}).  The procedure is illustrated in
Fig.~\ref{fig:3}. We defined two circular regions: a cluster field
containing G1 and a reference field nearby on the same CCD frame,
covering the same area of the cluster field. The radius of each
circular region is equal to 33.5\,arcsec ($\sim 1.5 \times r_{t}$,
\citealt{2007MNRAS.376.1621M}).  The stars of cluster and reference
fields are showed in the $m_{\rm F606W}$ versus $m_{\rm F438W}-m_{\rm
  F606W}$ CMDs of top-left and top-right panels of Fig.~\ref{fig:3},
respectively.  For each star $i$ in the reference field, we calculated
the distance
\begin{equation*}
\resizebox{0.48\textwidth}{!}{$d_i = \sqrt{(C^{}_{\rm F438W,F606W,cf}-C^i_{\rm F438W,F606W,rf})^2+(m_{\rm F606W,cf}-m^i_{\rm F606W,rf})^2}$}
\end{equation*}
where $C_{\rm F438W,F606W,cf}$ and $C^i_{\rm F438W,F606W,rf}$ are the
$m_{\rm F438W}-m_{\rm F606W}$ colours of the stars in the cluster and
reference fields, respectively. We flagged the closest star in the G1
field as a candidate to be subtracted. We associated to these flagged
stars a random number $0 \le r_i \le 1$ and we subtracted from the
cluster field CMD the candidates with $r_i<c^i_{rf}/c^i_{cf}$, where
$c^i_{rf}$ and $c^i_{cf}$ are the completeness of the star $i$ in the
reference field and of the closest star in the cluster field. Bottom
panels of Fig.~\ref{fig:3} show the result of the decontamination: left
panel show the $m_{\rm F606W}$ versus $m_{\rm F438W}-m_{\rm F606W}$
CMD of cluster field stars after the decontamination, while right
panel shows the stars subtracted from the cluster field CMD. Note the
similarity of G1 and field. Field is dominated by M31 stars whose CMD
is similar to G1.

\begin{figure*}
\includegraphics[width=\textwidth, bb=25 182 588 716]{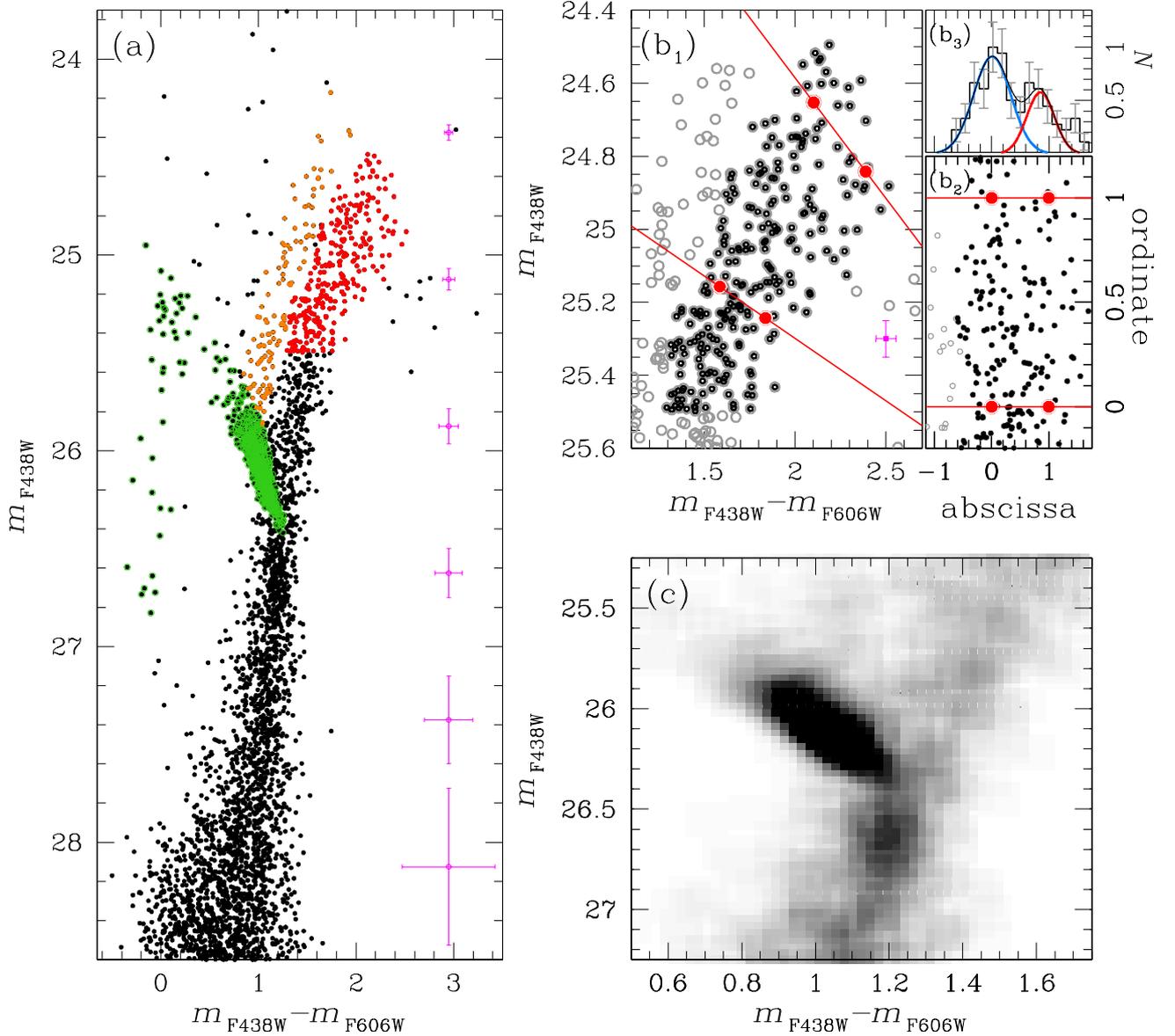}
\caption{Overview of G1 features. Panel (a) shows the $m_{\rm F438W}$
  versus $m_{\rm F438W}-m_{\rm F606W}$ CMD: RGB, HB, and AGB stars are
  highlighted in red, green, and orange, respectively. In magenta the
  observational errors in bins of 0.75 F438W magnitude.  Panel (b$_1$)
  is a zoom-in around the upper RGB region, where the sequence is
  splitted in two; in magenta the median observational error. Panel
  (b$_2$) shows the verticalized RGB sequence; the 'abscissa'
  distribution of the verticalized sequence is showed in panel
  (b$_3$). A bi-Gaussian function is fitted to the abscissa
  distribution. Panel (c) illustrates the Hess diagram around the CMD
  region containing the red clump and the RGB-bump.  \label{fig:4}}
\end{figure*}

\begin{figure*}
\includegraphics[width=\textwidth, bb=19 472 552 693]{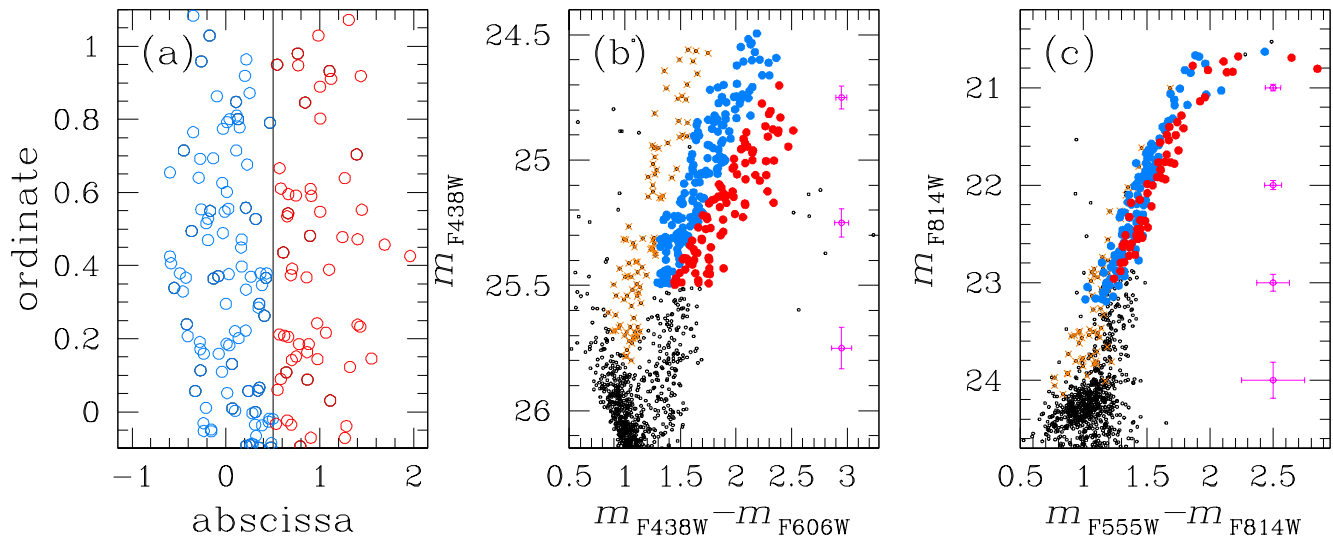}
\caption{Panel (a) shows the verticalized RGB sequence:
  RGB\,a (red) and RGB\,b (blue) stars are divided by the line at
  'abscissa'$=0.5$. Panels (b) and (c) show the RGB\,a (red) and
  RGB\,b (blue) stars in the $m_{\rm F438W}$ versus $m_{\rm
    F438W}-m_{\rm F606W}$ and $m_{\rm F814W}$ versus $m_{\rm
    F555W}-m_{\rm F814W}$ CMDs; AGB stars are plotted in
  orange. \label{fig:5}}
\end{figure*}

\begin{figure*}
\includegraphics[width=0.85\textwidth, bb=20 250 458 716]{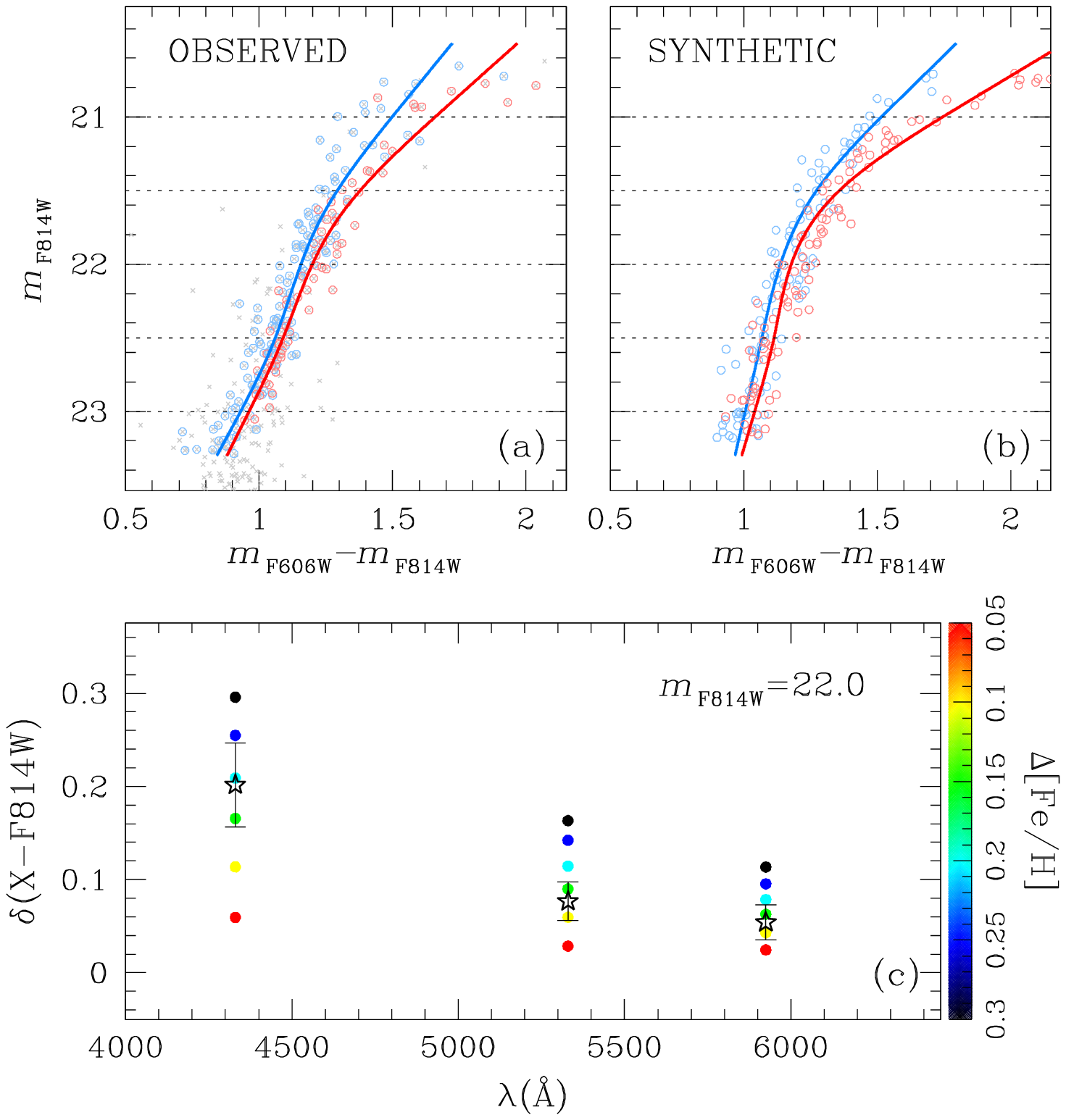}
\caption{Procedure adopted to derive $\Delta$[Fe/H] between RGB\,a and
  RGB\,b. Panel (a) shows a zoom of the $m_{\rm F814W}$ versus $m_{\rm
    F606W}-m_{\rm F814W}$ CMD around the upper part of the RGB. In red
  and blue the fiducial lines for RGB\,a and RGB\,b,
  respectively. Panel (b) shows the synthetic CMD in the same filters,
  obtained using the models with
  [Fe/H]=$-0.85$ (blue fiducial line) and
   [Fe/H]=$-0.70$ (red fiducial
  line). Panel (c) shows the comparison between the colour difference
  $\delta({\rm X-F814W})^{\rm obs}$ (black stars) obtained from
  observed fiducial lines and the colour differences $\delta({\rm
    X-F814W})^{\rm synth}$ (points colour-coded as in the bar on the
  right) for different $\Delta$[Fe/H] at $m_{\rm F814W}^{\rm
    cut}=22.0$.  \label{fig:6}}
\end{figure*}

\section{The CMD of G1}
\label{sec:cmd}

Left panel of Fig.~\ref{fig:2} shows the $m_{\rm F606W}$ versus $m_{\rm
  F438W}-m_{\rm F606W}$ CMD of all the well measured stars in the
field containing G1.

In panel (a) of Fig.~\ref{fig:4} we plot the decontaminated $m_{\rm
  F438W}$ versus $m_{\rm F438W}-m_{\rm F606W}$ CMD for all the well
measured sources of G1. We highlighted possible asymptotic giant
branch (AGB), RGB, and HB stars in orange, red, and green,
respectively. In the following we analyse the peculiarities of these
sequences and we compare the main properties of the CMD of G1 with
what observed in other massive Galactic GCs.

\subsection{The upper part of the RGB}
A visual inspection of the bright part of the $m_{\rm F438W}$ versus
$m_{\rm F438W}-m_{\rm F606W}$ CMD (panel (a) of Fig.~\ref{fig:4})
reveals that the upper part of the RGB sequence is broadened in
colour. The broadening becomes evident if we compare this sequence
with the average observational errors in colour (magenta crosses
\footnote{Error bars are obtained averaging the observational
    errors (the \texttt{RMS} parameter) in bins of 0.75 F438W
    magnitude.}) in the same range of magnitudes.  Also note that the
  fainter (and therefore affected by larger errors) blue HB is
  narrower than the brighter RGB, strengthening that the RGB
  broadening is intrinsic. This broadening was already observed by
  \citet{2001AJ....122..830M}, and was associated to a spread in
  metallicity among RGB stars.

Panel (b$_1$) of Fig.~\ref{fig:4} shows that the upper part of the RGB
of G1 splits in two sequences. We verticalized the RGB using the same
procedure adopted by \citet[see also \citealt{2009A&A...503..755M} for
  details]{2013ApJ...765...32B}. As a result of the verticalization
process, for each star we have a transformed colour and magnitude in a
new plane 'ordinate' versus 'abscissa' (panel (b$_2$)). The
distribution of the transformed colours of the RGB stars is shown in
panel (b$_3$): the distribution is bimodal, with two clear peaks. We
fitted the histogram with a bi-Gaussian function to derive the
fraction of stars belonging to the two sequences. We found that the
redder RGB (named RGB\,a) contains 32$\pm$2\,\% of the RGB stars,
while the RGB\,b, that is on the blue side of the RGB, contains
68$\pm$2\,\% of the RGB stars.

In Fig.~\ref{fig:5} we demonstrate that the bimodality of the upper
part of the RGB is a real feature: panel (a) shows the verticalized
RGB sequence as obtained in panels (b) of Fig.~\ref{fig:4}. We divided
the verticalized sequence in two groups, on the basis of the gap in
the RGB: the stars having 'abscissa'$<0.5$ (in blue, RGB\,b) and the
stars with 'abscissa'$>0.5$ (in red, RGB\,a). To investigate if the
separation of the two sequences is real, we used the approach adopted
in \citet{2015A&A...573A..70N,2018MNRAS.tmp..706N}: if the colour
spread is due to photometric errors, a star of the RGB\,a sequence
that is redder than the RGB\,b sequence in the $m_{\rm F438W}$ versus
$m_{\rm F438W}-m_{F606W}$ CMD (panel (b) of Fig.~\ref{fig:5}), should
have the same probability of being either red or blue in CMDs obtained
with other combinations of filters. The fact that in the totally
independent $m_{\rm F814W}$ versus $m_{\rm F555W}-m_{F814W}$ CMD
(panel (c) of Fig.~\ref{fig:5}) RGB\,a and RGB\,b form two well-defined
sequences demonstrates that the stars belonging to these two groups
have indeed different properties.

\subsection{Metallicity of G1 and spread of the RGB}
\label{subsec:met}
In literature there are many estimates of the metallicity for G1,
obtained using photometry (see, e.g.,
\citealt{2001AJ....121.2597S,2001AJ....122..830M,2003A&A...405..867B,2005AJ....129.2670R}),
integrated spectra (e.g., \citealt{1991ApJ...370..495H}), and
spectro-photometric Lick indices (\citealt{2009A&A...508.1285G}).
\citet{1991ApJ...370..495H} obtained [Fe/H]=$-1.08\pm0.09$ using
integrated spectra; by the analysis of the CMDs,
\citet{2001AJ....122..830M}, \citet{2001AJ....121.2597S}, and
\citet{2003A&A...405..867B} found [Fe/H]=$-0.95\pm0.09$,
      [Fe/H]=$-1.22\pm0.43$, and [Fe/H]=$-0.82\pm0.26$ respectively;
      from spectro-photometric Lick indices
      \citet{2009A&A...508.1285G} measured [Fe/H]=$-0.73\pm0.15$;
      on average the metallicity of G1 stars is $\sim
      -0.95$.

As already discussed in the previous section, the RGB
of G1 is strongly widened. \citet{2001AJ....122..830M} attributed this
spread to variations of [Fe/H] among RGB stars of G1.  In this section
we investigate if this hypothesis is valid, comparing the observed
split of the upper part of the RGB with theoretical models.

The procedure adopted is illustrated in Fig.~\ref{fig:6}. First, we
derived for the observed RGB\,a and RGB\,b the fiducial lines. We
derived the fiducial lines in the $m_{\rm F814W}$ versus $m_{\rm
  X}-m_{\rm F814W}$ CMDs, where X=F438W,F555W,F606W.  The procedure to
obtain a fiducial line is based on the naive estimator: we divided
the RGB\,a (RGB\,b) sequence in intervals of $\delta=1$\,F814W
magnitudes. On these intervals we defined a grid of $N$ points
separated by steps of width $\delta/10$ and we calculated the median
colour and magnitude within the interval $m^i_{\rm F814W}<m_{\rm
  F814W}<m^i_{\rm F814W}+\delta$, with $i=1,...,N$; we interpolated
these median points with a spline. An example of
fiducial lines is shown in panel (a) of Fig.~\ref{fig:6}, where the
red line is for RGB\,a and the blue line is for RGB\,b.

We obtained synthetic fiducial lines as follows: we considered a 12
Gyr $\alpha$-enhanced BaSTI isochrone
  (\citealt{2004ApJ...612..168P,2006ApJ...642..797P,2009ApJ...697..275P}) with a given
[Fe/H], and we obtained a first guess synthetic CMD interpolating the
isochrone on a vector of random $m_{\rm F814W}$ magnitudes, assuming a
flat luminosity distribution. We broadened this synthetic CMD by
adding to the colour of each synthetic star a random Gaussian noise,
with a dispersion equal to the average colour error measured for upper
RGB stars. Each synthetic CMD is composed by 10\,000 synthetic
stars. For each synthetic CMD we extracted synthetic fiducial lines
using the same approach adopted for the observed CMD. Panel (b) of
Fig.~\ref{fig:6} shows the synthetic CMDs{\footnote{For ease reading
    we plotted only 1\% of the synthetic stars} and the synthetic
  fiducial lines for two isochrones with 
  [Fe/H]=$-0.85$ (blue) and 
  [Fe/H]=$-0.70$ (red).

For each $m_{\rm F814W}$ versus $m_{\rm X}-m_{\rm F814W}$ observed
CMD, we computed the colour difference $\delta({\rm X-F814W})^{\rm obs}$
between  RGB\,a and RGB\,b fiducial lines, at 5 F814W magnitudes,
$m_{\rm F814W}^{\rm cut}=$21.0, 21.5, 22.0, 22.5, 23.

We chose the model [Fe/H]=$-0.70$ as
reference for RGB\,a \footnote{We found that
  the isochrone with [Fe/H]$=-0.70$ well fit the RGB\,a sequence in
  all the $m_{\rm F814W}$ versus $m_{\rm Y}-m_{\rm F814W}$ CMDs, where
  Y=F336W, F438W, F555W, and F606W} and, for each $m_{\rm F814W}^{\rm
  cut}$, we computed the colour difference $\delta({\rm X-F814W})^{\rm
  synt}$ between the synthetic fiducial line for
[Fe/H]=$-0.70$ and the fiducial lines for
$-1.00<$[Fe/H]$<-0.75$}, i.e.,
for $0.05<\Delta$[Fe/H]$<0.30$.

For each $m_{\rm F814W}^{\rm cut}$, we compare $\delta({\rm
  X-F814W})^{\rm obs}$ and $\delta({\rm X-F814W})^{\rm synt}$ to
search for the best $\Delta$[Fe/H] that reproduce the observed colour
difference in all the bands. Panel (c) of Fig.~\ref{fig:6} shows,
coloured from blue to red, the colour differences obtained from the
models for $m_{\rm F814W}^{\rm cut}=22.0$; black stars are the
observed points at the same magnitude level. In Table~\ref{tab2} we
listed the $\Delta$[Fe/H] obtained for each $m_{\rm F814W}^{\rm
  cut}$. We computed the weighted average of all the $\Delta$[Fe/H],
using as weight $1/\sigma_{\rm \Delta [Fe/H]}$, and we did so
  for three different cases: (i) the two populations have the same
  helium and C+N+O abundance (Scenario 1); (ii) the two populations
  share the same C+N+O content, but RGB\,a is populated by stars
  having primordial helium, while RGB\,b stars have $Y=0.30$ (Scenario
  2); (iii) the two populations have the same helium content, but
  RGB\,b is populated by stars C+N+O-enhanced (Scenario 3). In the
  latter case, the C+N+O content is about a factor of two larger than
  the CNO sum of the mixture used for the standard $\alpha$-enhanced
  case (\citealt{2009ApJ...697..275P}).  Because of the uncertainties
  on the bolometric corrections in blue and UV bands for the
  isochrones C+N+O-enhanced, for Scenario 3 we excluded the
  $\delta({\rm F438W-F814W})$ point in the computation of
  $\Delta$[Fe/H]. For the Scenario 1 we found that the split of the
  RGB is well reproduced with a mean metallicity difference of
  $\Delta$[Fe/H]$=0.15\pm0.03$, while in case of Scenario 2 and 3 we
  found $\Delta$[Fe/H]$=0.12\pm0.03$ and $\Delta$[Fe/H]$=0.14\pm0.04$,
  respectively.
  
We want to highlight that a similar behaviour of the RGB could also be
reproduced by  a combination of variations of
  metallicity, helium, and C+N+O content (see discussion in
  Sect.~\ref{sec:n6388}).

\begin{table}
  \begin{center}
  \caption{$\Delta$[Fe/H] values obtained for each $m_{\rm F814W}^{\rm cut}$}
  \label{tab2}
  \begin{tabular}{rrrr}
\hline
\multicolumn{1}{c}{} & \multicolumn{3}{c}{$\Delta$[Fe/H]} \\
{ $m^{\rm CUT}_{\rm F814W}$} & {Scenario 1} & {Scenario 2} & {Scenario 3} \\
\hline   
21.0 & 0.12  $\pm$ 0.04 &  0.10  $\pm$ 0.05  & 0.18  $\pm$ 0.05 \\
21.5 & 0.18  $\pm$ 0.06 &  0.13  $\pm$ 0.06  & 0.18  $\pm$ 0.07 \\
22.0 & 0.17  $\pm$ 0.05 &  0.12  $\pm$ 0.04  & 0.15  $\pm$ 0.05 \\
22.5 & 0.16  $\pm$ 0.06 &  0.11  $\pm$ 0.05  & 0.11  $\pm$ 0.05 \\
23.0 & 0.17  $\pm$ 0.06 &  0.12  $\pm$ 0.06  & 0.10  $\pm$ 0.05 \\
average     & 0.15  $\pm$ 0.03 & 0.12 $\pm$ 0.03 & 0.14  $\pm$ 0.04 \\
\hline
\end{tabular}

  \end{center}
\end{table}

\subsection{The HB of G1}
\label{sec:hb}
We know that the metallicity is the parameter that mainly determine
the morphology of the HB of a GC. However, some Galactic GCs, despite
they have the same metallicity, present different HB morphologies.
This is the so-called ``2nd-parameter'' problem. Many parameters have
been suggested as candidates to explain the peculiarities observed for
some Galactic GCs. Recently, \citet{2014ApJ...785...21M} have
demonstrated that the extension of the HB correlates with the mass of
the hosting GC. \citet{2018MNRAS.481.5098M} show that the
morphology of the HB is also correlated with the internal helium
variations among the different populations hosted by GCs,
demonstrating that HB morphology is strictly related to the presence
of multiple populations (see also, e.g., \citealt{2002A&A...395...69D,2011MNRAS.410..694D,2013MNRAS.430..459D}).
The HB morphology of M31 GCs and the effects of the second parameter
have been analysed in many works in literature (see, e.g.,
\citealt{2005AJ....129.2670R,2012A&A...546A..31P}).

Already \citet{1996AJ....111..768R} discovered that the HB of G1 is
very populated on the red side, and that a few stars populate the blue
side (see also \citealt{2013AAS...22121304R}). In this paper we
confirm that the HB of G1 is composed by a very populated red clump
and we show for the first time the evidence of a very extended blue HB
(green points in panel (a) of Fig.~\ref{fig:4}), more than expected and
observed until now, even though it is an intermediate metallicity GC
([Fe/H]$\sim -0.95$).

Taking into account of the completeness of our sample, we found that
the red clump and the blue side of HB of G1 contain $\sim 85$\% and
$\sim 15$\% of HB stars, respectively. It has been demonstrated that
Galactic GCs that show HB morphologies similar to that of G1 also host
a small fraction of very high helium-enhanced stars (see, e.g.,
\citealt{2007A&A...463..949C,2007A&A...474..105B,2013ApJ...765...32B,2014ApJ...785...21M,2017MNRAS.465.1046T})
that populate the blue HB. We suggest that also G1 might host a small
fraction ($\sim 15$\,\%) of highly helium-enhanced stars, that evolve
into the blue side of the HB.

\section{Comparison between G1 and Galactic GCs}
\label{sec:compare}
In this section we compare the massive GC G1 in M31 with Galactic GCs
of similar mass. We compared G1 with two
Galactic GCs, NGC\,6388, and NGC\,6441, using the
HUGS\footnote{\url{https://archive.stsci.edu/prepds/hugs/}} catalogues
published by \citet{2018MNRAS.481.3382N}.

\begin{figure}
\includegraphics[width=0.5\textwidth, bb=20 330 315 716]{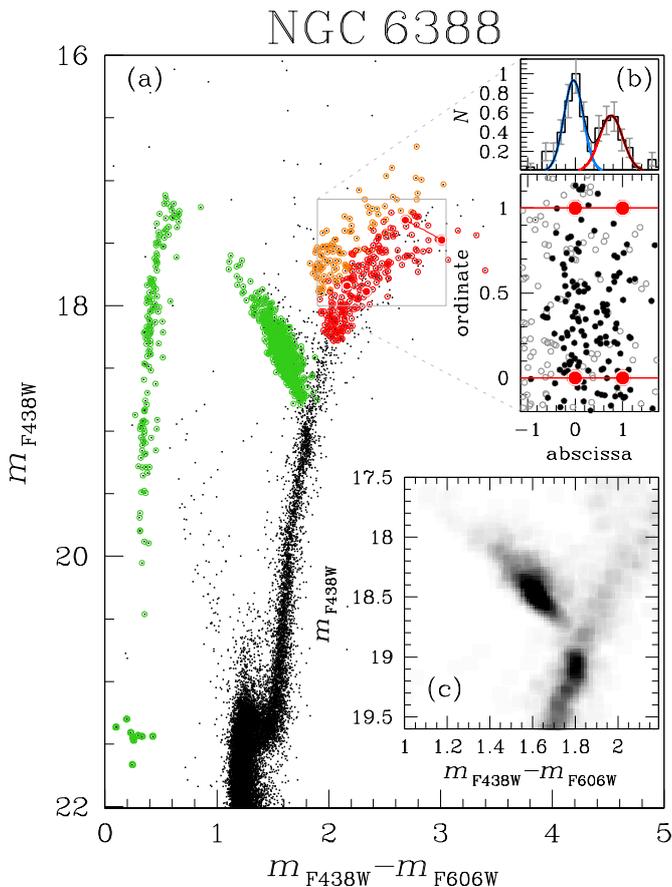}
\caption{Overview of NGC\,6388 features. Panel (a) shows the $m_{\rm
    F438W}$ versus $m_{\rm F438W}-m_{\rm F606W}$ CMD: HB, RGB and AGB
  stars are highlighted in green, red and orange, respectively.
  Panels (b) shows the verticalized RGB sequence using the stars of the grey box of
  panel (a); the 'abscissa' distribution of the verticalized sequence
  is shown in the upper panel. A bi-Gaussian function is fitted to
  the abscissa distribution. Panel (c) illustrates the Hess diagram
  around the CMD region containing the red clump and the RGB bump.
    \label{fig:7}}
\end{figure}

\begin{figure*}
\includegraphics[width=1.0\textwidth, bb=20 320 583  691]{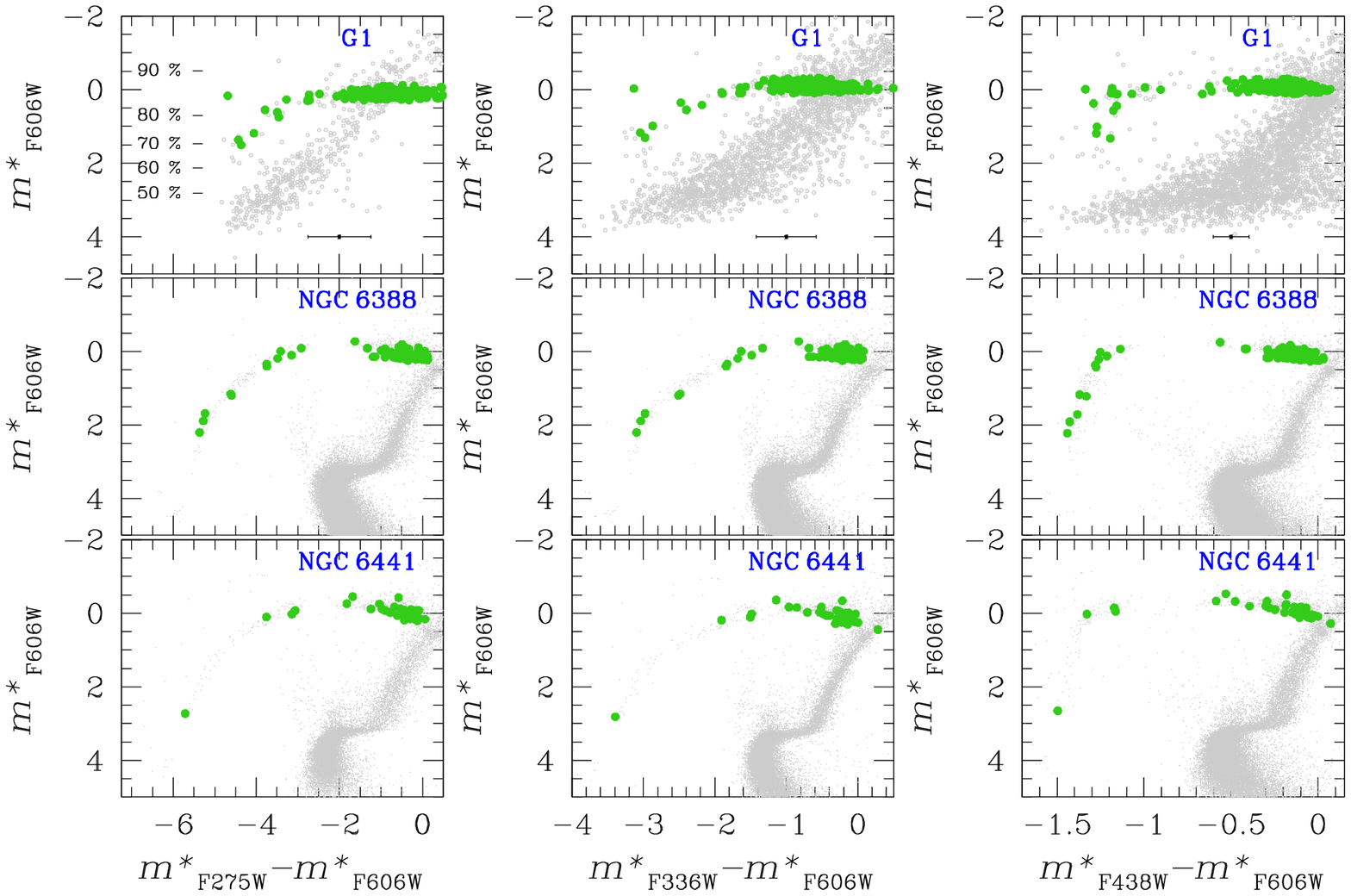}
\caption{Transformed $m^*_{\rm F606W}$ versus $m^*_{\rm X}-m_{\rm
    F606W}$ CMDs, with X=F275W, F336W, and F438W, for the HB stars of
  G1 (first row), NGC\,6388 (second row), NGC\,6441 (third row). The
  left-hand panels are the $m^*_{\rm F606W}$ versus $m^*_{\rm
    F275W}-m_{\rm F606W}$ CMDs, the middle panels are the $m^*_{\rm
    F606W}$ versus $m^*_{\rm F336W}-m_{\rm F606W}$ CMDs, while the
  right-hand panels show the $m^*_{\rm F606W}$ versus $m^*_{\rm
    F438W}-m_{\rm F606W}$ CMDs of the GCs listed above. The green
  points are the HB stars located between $2.5\times r_{h}$ and $4.5
  \times r_{h}$ from the cluster centers. Top-left panel reports the
  completeness of G1 catalogue at different magnitude levels.
 \label{fig:8}}
\end{figure*}

\subsection{G1 versus NGC\,6388}
\label{sec:n6388}

The GC NGC\,6388 is a massive, metal-rich ([Fe/H]=$-0.45\pm0.04$,
\citealt{2009A&A...508..695C}) Galactic globular cluster, with an
anomalously extended blue HB  (for its metallicity).

From a visual inspection of the $m_{\rm F438W}$ versus $m_{\rm
  F438W}-m_{\rm F606W}$ CMD of G1 (panel (a) of Fig.~\ref{fig:4}) and
NGC\,6388 (panel (a) of Fig.~\ref{fig:7}) we note many common features
including the well populated red clump, the extended blue HB, and the
splitted RGB.

As already demonstrated by \citet{2013ApJ...765...32B}, the upper part
of the RGB of NGC\,6388 splits in two sequences in the $m_{\rm F438W}$
versus $m_{\rm F438W}-m_{\rm F606W}$ CMD. Panels (b) of
Fig.~\ref{fig:7} shows the verticalized RGB sequence (bottom) and the
histogram of the ``abscissa'' distribution (top) for NGC\,6388,
obtained as in Sect.~\ref{sec:cmd}. By least-squares fitting a
bi-Gaussian function, we found that the RGB on the red side contains
$42\pm2$\,\% of RGB stars, while the remaining $58\pm3$\,\% stars are
on the blue side. This result is very similar to what we found for G1.

In Sect.~\ref{subsec:met} we have tested the hypothesis that the RGB
split observed for G1 is due to a metallicity variation of 
$\sim 0.15$ dex between the two populations. For NGC\,6388,
\citet{2007A&A...464..967C} found no intrinsic metallicity spread from
spectroscopic measurements of RGB stars. A possible explanation of the
RGB split could be a variation of the C+N+O content between the two
populations, that implies also an SGB split in the optical CMDs. The
SGB split for NGC\,6388 was noticed by \citet{2013ApJ...765...32B}.
Unfortunately, on the dataset used in this work, we are not able to
distinguish between metallicity or C+N+O variations as cause of the
RGB split observed for G1, because the SGB stars of G1 are too faint
to be measured.

The HB morphologies of G1 and NGC\,6388 are also very similar (in
green in Figs.~\ref{fig:4} and \ref{fig:7}). Both HBs present a
populated red clump, containing $\sim 85$-$90$\,\% of HB stars and well
spread in magnitude. \citet{2017MNRAS.465.1046T} showed that the red
side of the HB of NGC\,6388 is compatible with the presence of
He-enriched stars ($\Delta Y \sim 0.08$).

The extended blue HB of G1 is very similar to that of NGC\,6388.  In
the case of NGC\,6388, \citet{2007A&A...474..105B} and \citet{2017MNRAS.465.1046T} showed that stars
on the blue side of the HB of NGC\,6388 are very helium-enriched, up
to $\Delta Y \sim 0.13$ for the bluest ones.

There are few
differences between the two clusters that must be taken into account in this analysis.
First of all, NGC\,6388 is more metal-rich ([Fe/H]=$-0.45\pm0.04$,
\citealt{2009A&A...508..695C}) than G1 ([Fe/H]$\sim -0.95$). Moreover, the present-day mass of NGC\,6388
($\sim 2\times10^6\,M_\odot$,
\citealt{2011ApJ...742...51B}), is almost one order of magnitude smaller
than G1 mass ($\sim 1.5\times10^7\,M_\odot$
\citealt{2001AJ....122..830M}). Finally, the position of the two
clusters in the host galaxy is totally different: NGC\,6388 is a bulge
Galactic GC, while G1 is projected at 40\,kpc in the halo of M31.

\subsection{Multi-band comparison of the HB morphologies}

The dataset used in this work allows us to analyse the extended HB of
G1 from the near UV to the visible.

To compare the HB morphology of G1 with that of other Galactic GCs,
for each $m_{\rm F606W}$ versus $c_{\rm X}=m_{\rm X}-m_{\rm F606W}$
CMD, where X=F275W, F336W, and F438W, we performed a translation of
the entire CMD as follows: we calculated the 96th percentile (on the
red side) of the HB colour distribution, $c_{\rm 96th, X}$, and the
associated magnitude, $m_{\rm 96th, F606W}$ Then, we defined a new
colour $c^*_{\rm X}=c_{\rm X}-c_{\rm 96th, X}$ and magnitude $m^*_{\rm
  F606W}=m_{\rm F606W}-m_{\rm 96th, F606W}$.

To compare the HBs of G1 and of the Galactic GCs NGC\,6388 and
NGC\,6441, avoiding the very low completeness of the G1 catalogue in the
center region, we selected only the HB stars between $2.5\times r_{h}$
and $4.5 \times r_{h}$ from the cluster centers, where $r_h$ is the
half-light radius ($r_h=$1.7, 31.2, and 34.2\,arcsec for G1,
NGC\,6388, and NGC\,6441, respectively,
\citealt{1996AJ....112.1487H,2007MNRAS.376.1621M}).

The first row of Fig.~\ref{fig:8} shows the multiband photometry of the
HB of G1 (green points). In the left panel we also report the completeness level.
We compared the HB morphology of G1 with that of two Galactic GCs that
are massive ($\gtrsim 10^6\,M_\odot$), that host stars highly enriched
in helium ($Y\sim 0.30-0.35$) and that have a populated red clump:
NGC\,6388 (2nd row of Fig.~\ref{fig:8}) and NGC\,6441 (3rd row of
Fig.~\ref{fig:8}). We highlighted in green their HB stars.

The GCs NGC\,6388 and NGC\,6441 are [Fe/H]-rich clusters ($\sim
-0.45$; \citealt{2009A&A...508..695C}) located in the bulge of the
Milky Way. Both clusters show a peculiar HB morphology: a very
populated red clump (85-90\% of HB stars) and an extended blue HB
(\citealt{1997ApJ...484L..25R}).  Even if these two clusters seem to
be twins, the {\it chromosome maps} of RGB stars
(\citealt{2017MNRAS.464.3636M}) are very different.

A comparison between the first, the second and the third row of
Fig.~\ref{fig:8} illustrates how the HB morphology of G1 is similar to
that of the two selected Galactic GCs in all the analysed filters: the
populated and extended red clumps, and the blue HB that covers about
the same colour ranges are common features to these three GCs.

We know that both the red clumps of NGC\,6388 and of NGC\,6441 host
stars with primordial helium content and helium enhancement by $\Delta
Y\sim 0.08-0.1$, while the blue HB is populated by stars that have
helium content $Y>0.35$ (\citealt{2017MNRAS.465.1046T}). It is
possible that a similar helium enhancement might plausibly explain the
extended HB of G1.

The GCs NGC\,6388 and NGC\,6441 host stars that form an extreme HB
(eHB, \citealt{2013ApJ...765...32B,2016ApJ...822...44B}, see also
Fig.~\ref{fig:7}). These stars are located mainly in the center of the
clusters. The cluster G1 does not show eHB stars: we attribute the
lack of such stars to the extreme crowding and consequent low
completeness at these magnitudes in the central regions of the
cluster.

\section{Summary}
\label{sec:summary}
In this work we presented a multi-band analysis of the CMDs of the GC
Mayall\,II (G1) located in the halo of M31. We used {\it HST} data
collected with filters that cover wavelengths from $\sim 250$\,nm to
$\sim 950$\,nm. This is the first time that G1 is observed in UV and blue filters.

The high accuracy photometry extracted from {\it HST} data with new,
advanced tools, has allowed us to extract for the first time a deep
CMD (magnitude limits $V\sim 28.5$, $\sim 3$ magnitudes below the HB
level) for this cluster. The $m_{\rm F606W}$ versus $m_{\rm
  F438W}-m_{\rm F606W}$ CMD shows new features never observed until
now: a wide, likely bifurcated RGB, an extended blue HB, and populated RGB bump, AGB
and red clump.

We show that the upper part of the RGB of G1 splits in two real
sequences in the $m_{\rm F438W}$ versus $m_{\rm F438W}-m_{\rm F606W}$
CMD, as for the Galactic GC NGC\,6388. We derive the fraction
of stars within each sequence and we found the the redder sequence
(RGB\,a) includes $\sim 32$\,\% of RGB stars, while the bluer
RGB\,b sequence contains the remaining $\sim 68$\,\%. 

We tested the hypothesis of [Fe/H]-variations to explain the RGB
split, comparing the observed fiducial lines of the two populations
with synthetic fiducial lines obtained using models. We found that the
RGB split could be reproduced assuming that the two populations have a
difference in [Fe/H]  of $\Delta$[Fe/H]$= 0.15 \pm 0.03$ if the
  populations share the same helium and C+N+O content,
  $\Delta$[Fe/H]$=0.12\pm 0.03$ if the two populations have different
  helium content and the same C+N+O content, and $\Delta$[Fe/H]$=
  0.14 \pm 0.04$ if the two populations share the same helium abundance but
  have different C+N+O content.

We compared the HB morphology of G1 with that of massive Galactic GCs
that present a populated red clump.  We found many similarities with
the HBs of NGC\,6388 and NGC\,6441: a populated red clump ($>85$\,\%
of HB stars) and an extended blue HB, despite these GCs are
[Fe/H]-rich. We know that the red clump and the blue HB of NGC\,6388
and NGC\,6441 are populated by stars helium enriched by $\Delta Y\sim
0.08$ and $\Delta Y \sim 0.13$, respectively. Given the similarity
between the HB morphologies of G1 and NGC\,6388/NGC\,6441, we can
conclude that also HB stars of G1 may be helium enriched.  An
additional proof is given by the correlation found by
\citet{2018MNRAS.481.5098M} for Galactic GCs between the maximum
internal helium variation and the mass of the cluster: more massive
clusters have larger internal maximum helium variation. Because G1 is
$\sim 3$ times more massive than $\omega$~Cen, the most massive GC in
the Milky Way ($\sim 4 \times 10^6 M_\odot$,
\citealt{2013MNRAS.429.1887D}),
we expected
that 
the maximum internal helium variation among G1 stars is
high, with $\Delta Y \gtrsim 0.1$.

All the results obtained in this work represent a proof that G1 hosts
multiple stellar populations, characterised by different chemical
properties. Moreover, some evidence tell us that G1 could belong to
the group of ``anomalous'' GCs, that host populations with different
C+N+O contents and/or [Fe/H]-content (as already suggested by
\citealt{2001AJ....122..830M}). In the Milky Way a significant
fraction of globular clusters belongs to the group of ``anomalous''
GCs ($\omega$\,Cen, M22, NGC\,5286, M54, M19, M2, NGC\,6934, etc.,
see, e.g., \citealt{1996ASPC...92..375N, 2010A&A...520A..95C,
  2011A&A...532A...8M, 2014MNRAS.441.3396Y, 2015MNRAS.450..815M,
  2015AJ....150...63J, 2017ApJ...836..168J, 2018ApJ...859...81M}), and
  the discovery of a GC in M31 with similar characteristics means that
  these kind of clusters are not a prerogative of our Galaxy.
  Unfortunately, the dataset used in this work is not sufficient to
  constraint the chemical properties of this GC and of its
  populations, and more multi-band photometric observations are
  required to trace the formation and evolution of G1.

\section*{Acknowledgements}
DN and GP acknowledge partial support by the Universita` degli Studi
di Padova Progetto di Ateneo BIRD178590. SC acknowledges support from
Premiale INAF ``MITIC'', from INFN (Iniziativa specifica TAsP), and
grant AYA2013-42781P from the Ministry of Economy and Competitiveness
of Spain. 




\bibliographystyle{mnras}
\bibliography{biblio}




\bsp	
\label{lastpage}
\end{document}